\documentclass[12pt, hidelinks]{article}
\usepackage[utf8]{inputenc}
\usepackage{amsmath, amssymb, url, float, graphicx, float}
\usepackage{algorithm, algpseudocode}
\usepackage{hyperref}
\usepackage[small,bf]{caption}
\usepackage{fullpage}

\newcommand{\da}{\Delta_\alpha}
\newcommand{\db}{\Delta_\beta}
\newcommand{\duni}{\Delta_\mathrm{UNI}}
\newcommand{\ra}{R_\alpha}
\newcommand{\rb}{R_\beta}
\newcommand{\runi}{R_\mathrm{UNI}}
\newcommand{\eps}{\varepsilon}

\newcommand{\downto}{\downarrow}
\newcommand{\upto}{\uparrow}

\newcommand{\ones}{\mathbf 1}
\newcommand{\reals}{{\mbox{\bf R}}}

\newcommand{\symm}{{\mbox{\bf S}}}  

\newcommand{\Expect}{\mathop{\bf E{}}}

\newcommand{\eg}{{\it e.g.}}
\newcommand{\ie}{{\it i.e.}}

\newcommand{\BEAS}{\begin{eqnarray*}}
\newcommand{\EEAS}{\end{eqnarray*}}
\newcommand{\BEA}{\begin{eqnarray}}
\newcommand{\EEA}{\end{eqnarray}}
\newcommand{\BEQ}{\begin{equation}}
\newcommand{\EEQ}{\end{equation}}
\newcommand{\BIT}{\begin{itemize}}
\newcommand{\EIT}{\end{itemize}}

\title{An analysis of Uniswap markets}
\author{Guillermo Angeris\footnote{\texttt{angeris@stanford.edu}} \and Hsien-Tang Kao\footnote{\texttt{\{hsien-tang, rei, tarun\}@gauntlet.network}} \and Rei Chiang\footnotemark[2]
\and Charlie Noyes\footnote{\texttt{charlie@paradigm.xyz}} \and Tarun Chitra\footnotemark[2]}
\date{November 2019}

\begin{document}
\maketitle

\begin{abstract}
Uniswap---and other constant product markets---appear to work well in practice despite their simplicity.
In this paper, we give a simple formal analysis of constant product markets and their generalizations, showing that, under some common conditions, these markets must closely track the reference market price.
We also show that Uniswap satisfies many other desirable properties and numerically demonstrate, via a large-scale agent-based simulation, that Uniswap is stable under a wide range of market conditions.
\end{abstract}

\section{Introduction}
Smart contract systems such as Ethereum~\cite{wood2014ethereum} and Tezos~\cite{goodman2014tezos} have allowed for the design and implementation of decentralized versions of traditional financial primitives.
The use of these primitives has grown dramatically in 2019: in January, less than \$20 million of Ethereum-based assets were utilized in these systems, a number which increased to \$750 million by December~\cite{murphy_2019}.
Decentralized financial primitives allow for censorship-resistant participation in a number of digital markets, expanding the reach of lending \cite{compound2019}, stable assets \cite{celo,team2017dai}, and exchanges \cite{uniswap,0x,bancor} past the conventional financial world.
In particular, a secure decentralized exchange for cryptocurrencies has been desired almost since the advent of Bitcoin as centralized exchanges, such as Mt. Gox~\cite{decker2014bitcoin}, Quadriga~\cite{shane2019crypto}, and Bitfinex~\cite{kaminska2016bitcoin}, have suffered catastrophic losses aggregating to billions of dollars' worth of depositors' funds.  

Historically, many different decentralized exchanges (DEXs) have been proposed using different market maker mechanisms, ranging from classic order book mechanisms~\cite{0x} to other, more complicated approaches with particular bonding curves~\cite{bancor}.
Yet, a simple but surprisingly effective market maker appears to be the constant product market maker used by Uniswap~\cite{uniswap}, likely the first and possibly the most popular implementation.
These markets provide a simple approach for trading between pairs of coins in a decentralized fashion, and have, in recent years, become a popular (and practically useful) alternative to other types of DEXs.

Additionally, other protocols (such as Celo~\cite{celo}) have used the idea of constant-product markets as a decentralized price oracle --- a contract that can be queried to find the relative price between two coins of interest.
These protocols make use the fact that, if the price indicated by the oracle differs from the true market price of the traded coins, an arbitrageur can always make a profit by trading with this oracle, implying that the price indicated by this oracle is likely close to the reference market price.

\paragraph{Comparison with other DEXs.}
While order book mechanisms are the dominant medium of exchange of electronic assets in traditional finance~\cite{wyart2008relation}, they are challenging to use within a smart contract environment.
The size of the state needed by an order book to represent the set of outstanding orders (\eg, passive liquidity) is large and extremely costly in the smart contract environment, where users must pay for space and compute power utilized~\cite{wood2014ethereum}.
Moreover, the matching logic for order books is often complicated as it must often support several different order types (such as icebergs, good-till-cancel, and stop-limit orders~\cite{wyart2008relation}).

In order to avoid the costly on-chain execution costs (paid to miners/validators of the smart contract by agents executing trades) a variety of designs for decentralized exchanges use the blockchain underlying a smart contract for settlement, while executing trades off-chain~\cite{0x,juliano2017dydx}.
These exchanges, however, have a number of drawbacks.
First, the complicated interaction between off-chain and on-chain mechanisms, coupled with the transaction ordering ambiguity inherent in blockchain-based systems, allows for front-running, which has been observed in practice~\cite{daianFlashBoysFrontrunning2019}.
Second, keeping the order book state in the hands of multiple participants (such as `relayers,' in 0x parlance~\cite{0x}) often leads to stale orders and latency arbitrage that is many orders of magnitude worse than that of high-frequency trading.
Finally, these exchanges have a more complicated threat model than that of a smart contract which does not have to interact with an exogenous, non-blockchain system to determine state transitions.
Due to this, users must often take extra security precautions when trading or potentially use complicated exit games to release their funds.

\paragraph{Automated market makers.} On the other hand, automated market makers (AMMs) have been studied extensively in algorithmic game theory, beginning with Hanson's logarithmic market scoring rule (LMSR)~\cite{hanson2003combinatorial}, often used in practice as an AMM for prediction markets.
Such AMMs are constructed by first having liquidity providers deposit assets in some fixed ratio to specify an initial distribution of beliefs over possible outcomes.
An AMM then provides a \emph{scoring rule} which specifies the cost of changing the distribution of beliefs from its current state to a new, desired state. This scoring rule incentivizes traders to report their true belief such that the expected value of adjusting the distribution is positive.
Since the state of the exchange depends only on the total amount of quantities deposited, the corresponding storage requirements are much smaller than those of traditional exchanges. Additionally, pricing a trade requires only a single function evaluation, as opposed to more complicated matching algorithms, as in the case of order books.

\paragraph{Bonding curves.} In general, LMSRs (and similar AMMs) are designed to predict the outcome of some set of (disjoint) events rather than predict a specific price. In many cases, the proposed mechanism cannot be directly used to price arbitrary assets without requiring a large state space and often suffers from other practical issues such as attracting liquidity~\cite{othman2013practical}.

To solve this problem, early Ethereum-based AMMs such as Bancor~\cite{bancor} moved to a second model for pricing assets: in this model, the function specifies the cost of an asset based on the total available supply (as, for example, its tokens are minted or burned), rather than specifying the cost of changing a given distribution. The function itself is called a bonding curve, and the resulting equilibrium price of the asset is then equal to the market price under certain conditions.

\paragraph{Uniswap.} Another possible model for pricing assets, first introduced by Uniswap~\cite{uniswap}, does not require the ability to change the supply of an asset in order to measure its price. Rather, an AMM holds some quantity of assets whose relative price we wish to measure in its reserves. The AMM then specifies a pricing function, which maps the quantities of the assets in reserves to their marginal price (with respect to any num\'eraire). Agents are then allowed to trade with this contract at the price specified by the pricing function, and this price is continually updated as its reserves change after each trade. For example, constant product markets such as Uniswap are specific cases in which the pricing function is exactly equal to the ratio of the reserves available to the contract, when no trading fees are taken (\S\ref{sec:opt-arb}).

While Uniswap, and its associated class of AMMs, is similar in spirit to bonding curve-based AMMs, we will distinguish them as a separate class of AMMs with a fairly distinct range of applicability. We will focus only on Uniswap-like AMMs---the constant product and constant mean AMMs---and leave the discussion of bonding curves and their theoretical properties for future research.

\paragraph{Summary.}
In this paper, we present optimal arbitrage actions and bounds under some simple reference market models for Uniswap's AMM model, which we call the \emph{constant product markets}, showing that Uniswap must closely track the reference market price under common market conditions.
We also show that a recent generalization of constant product markets, the \emph{constant mean markets}, first proposed in~\cite{balancer}, have nearly identical theoretical properties and may be of future interest.
Finally, we run a large-scale simulation of agents interacting with the Uniswap contract over a wide range of market parameters, suggesting that the system may be stable under a variety of market conditions.
These results help us conclude that Uniswap serves as a censorship resistant price oracle for smart contracts, provided that there exists an external reference market with sufficient liquidity.  

\section{Constant product markets}\label{sec:cpm}
\label{sec:const-prod}
A \emph{constant product market}~\cite{uniswap} is a market for trading coins of type $\alpha$ for coins of type $\beta$ (and vice versa). This market has reserves $\ra > 0$ and $\rb > 0$, constant product $k = \ra\rb$, and percentage fee $(1-\gamma)$. A transaction in this market, trading $\db > 0$ coins $\beta$ for $\da > 0$ coins $\alpha$, must satisfy
\begin{equation}
\label{eq:const-prod}
(\ra - \da)(\rb + \gamma \db) = k.
\end{equation}
After each transaction, the reserves are updated in the following way: $\ra \mapsto \ra - \da$, $\rb \mapsto \rb + \db$, and $k \mapsto (\ra - \da)(\rb + \db)$. We will always require that $\ra, \rb > 0$, such that any trade that results in a nonpositive reserve is never fulfilled (equivalently, we say that such a trade has infinite cost).

The name `constant product market' comes from the fact that, when the fee is zero (\ie, $\gamma = 1$), any trade $\db$ to $\da$ must change the reserves in such a way that the product $\ra\rb$ remains equal to the constant $k$.

In this section, we derive bounds on the marginal price of the market relative to a reference market and show that this market maker has other desirable properties.

\subsection{Optimal arbitrage in Uniswap}
\label{sec:opt-arb}
In the optimal arbitrage problem, we have two coins, $\alpha$ and $\beta$, which we can trade either with a reference market or a Uniswap contract. In this problem, we seek to maximize the profit made from trading, say, some amount of loaned coin, $\db$ of coin $\beta$ to some amount $\da$ of coin $\alpha$ via the Uniswap market. We then trade back the received $\da$ for $\db'$ and pay back the loan $\db$ to receive profit $\db' - \db$.

If our profit is positive (that is, if $\db' - \db > 0$), then we say that there is an \emph{arbitrage opportunity}, since we have made money `for free' (\ie, by only trading coins within different markets) without taking on any risk. The optimal arbitrage problem then asks: what is the maximum profit that can be made by this scheme?

In the infinitely liquid market case, \ie, in the case that $\db' = m_p \da$ (where $m_p$ is the reference market price of coin $\alpha$), we can phrase the optimal arbitrage problem as the following optimization problem:
\begin{equation} \label{eq:main}
\begin{aligned}
& \text{maximize} & & m_p\da - \db\\
& \text{subject to} & & \da, \db \ge 0\\
&&& (R_\alpha - \da)(R_\beta + \gamma \db)=k,
\end{aligned}
\end{equation}
with optimization variables $\da \in \reals$ and $\db \in \reals$, constrained to be nonnegative. Here, $(1 - \gamma)$ is the Uniswap exchange fee, and the constraint is the definition of a constant product market~\eqref{eq:const-prod}.

When written in this form, problem~\eqref{eq:main} is not obviously convex, though we show in appendix~\ref{app:Uniswap-convex} that it can easily be written in a convex form and then derive an analytical form for the optimal trade amounts $\da^\star$ and $\db^\star$.

\paragraph{No-arbitrage conditions.}
Assuming that the no-arbitrage condition is satisfied (which often approximately holds in practice; see~\cite[\S1.2]{financial}), we can show that the Uniswap market price deviates from the market price by at most a factor of $\gamma$.

The marginal price of coin $\alpha$ in Uniswap is defined as the price of an infinitesimally small trade. This price can be found by differentiating the constant-product market formula:
\[
\frac{d}{d\da}\left((\ra -\da)(\rb + \gamma \db)\right) = 0 \implies \frac{d\db}{d\da}\bigg|_{\da = 0} = \frac{1}{\gamma}\frac{\rb}{\ra} = \gamma^{-1} m_u,
\]
where we have written $m_u = \rb / \ra$ for the Uniswap price of coin $\alpha$ without the fee. 

We can always make a nonzero profit if the Uniswap marginal price of $\alpha$, $\gamma^{-1}m_u$ is smaller than the market marginal price $m_p$, by performing a small enough trade. Assuming there is no arbitrage, this means we must have
\[
m_u \ge \gamma m_p.
\]
Similarly, by swapping $\alpha$ for $\beta$ and combining the resulting statements we get the following bounds on the Uniswap market price, relative to the true market price, $m_p$:
\begin{equation}\label{eq:no-arb}
\gamma m_p \le m_u \le \gamma^{-1}m_p,
\end{equation}
assuming no-arbitrage conditions.

This suggests that, in practice, the larger the trade fees are, the larger the gap between the true market price and the Uniswap market price may be. For example, in the case that the trade fee $\tau = 1 - \gamma$ is small, bound~\eqref{eq:no-arb} is approximately equivalent to
\[
(1-\tau) m_p \le m_u \le (1+\tau)m_p.
\]

While we derived these conditions using a simple argument, we can also easily derive them by analyzing problem~\eqref{eq:main} and noting that its optimal value is nonzero if, and only if, there is an arbitrage opportunity. For more details, see appendix~\ref{app:Uniswap-convex}.

\subsection{Extensions of the optimal arbitrage problem}
\label{sec:extensions}
There are several natural extensions to the optimal problem~\eqref{eq:main}, which retain most of its useful properties.

\paragraph{Risk models.}
There are many factors which could potentially cause an agent to fail to close an arbitrage opportunity, including noisy information, front-running~\cite{daianFlashBoysFrontrunning2019}, and delay in trades---the latter of which is quite common in distributed platforms. In particular, it may not be desirable for an agent to perform a trade with the exact solution of problem~\eqref{eq:main}.

In this case, we can add an additional term penalizing large trades: in other words, we can assign some `cost of risk' which can be any convex function $f:\reals_+^2 \to \reals$ which is nondecreasing in its second argument. This gives a problem of the form:
\begin{equation} \label{eq:agent-arb}
\begin{aligned}
& \underset{\da, \db}{\text{maximize}} & & m_p\da - \db - f(\da, \db)\\
& \text{subject to} & & (R_\alpha - \da)(R_\beta + \gamma \db)= k\\
&&& \da, \db \ge 0.
\end{aligned}
\end{equation}
It should be noted that problem~\eqref{eq:agent-arb} can be written in a convex way and is the problem we use to simulate arbitrageurs in our agent simulation (see~\S\ref{sec:agents} for more details).

One example of a risk function $f$ could be a model of the resulting changes in market price due to the arbitrage trade. A common (and simple) model for the marginal market price $m: \reals_+ \to \reals_+$ after a trade of size $0 \le \da \le \eta^{-1/\xi}$ is that the resulting price is
\[
m(\da) = m_p\left(1 - \eta \da^\xi\right),
\]
with any $\eta \ge 0$, $\xi > 0$. The resulting risk function, $f(\da) = \int_0^{\da}(m_p - m(t))\,dt$, is convex. See appendix~\ref{app:Uniswap-extensions} for more details and extensions.

Additionally, in this model, unlike in problem~\eqref{eq:agent-arb}, an arbitrageur is not guaranteed to reach the no-arbitrage condition in a single round if the penalty function does not reflect the true underlying market movement.

\subsection{Other conditions}
\label{sec:other-properties}
In this section, we show a few basic and useful properties which Uniswap (and other constant-product markets) satisfy.

\paragraph{Increasing product constant.}
For every trade, the product constant $k$ is nondecreasing (and strictly increasing if $\gamma < 1$). Let $\ra^t > 0$, $\rb^t > 0$, and $k^t > 0$ be the reserve of $\alpha$, $\beta$, and the constant product at trade $t$. If trade $t+1$ sells $\db$ coins $\beta$ for $\da$ coins $\alpha$, we have (by definition, see~\eqref{eq:const-prod}):
\begin{equation}\label{eq:increasing-constant}
k^t = (\ra^t - \da)(\rb^t + \gamma \db) \le (\ra^t - \da)(\rb^t + \db) = k^{t+1},
\end{equation}
where inequality always holds strictly except when $\gamma = 1$. So, if $\gamma < 1$, we have that $k^t < k^{t+1}$, always.

\paragraph{Splitting trades is more expensive.} Whenever $\gamma < 1$, any agent agent trading some amount $\da > 0$ for some output $\db$ and immediately trading another amount $\da' > 0$ for $\db'$ will always have smaller output than if the agent had traded $\da + \da'$ for some output $\db^\mathrm{tot}$ all at once (this property is also sometimes called \emph{path-dependence}~\cite{othman2013practical}). This can easily be proven, though it is mostly an exercise in algebra. We outline the steps in appendix~\ref{app:larger-incentives}.

This fact implies that, in the case where the reference market is infinitely liquid (as in~\S\ref{sec:opt-arb}), an arbitrage agent who considers a strategy several steps into the future cannot do better than simply solving~\eqref{eq:main} and executing the corresponding trade.

\paragraph{No-depletion property.}
It turns out to be impossible to fully deplete Uniswap of all coins only by trading $\alpha$ and $\beta$, even if the attacker has an unbounded amount of coins. We can easily show that the total reserve is always bounded from below. This follows immediately from applying the arithmetic-geometric mean inequality~\cite[\S3.1.9]{cvxbook}:
\[
2\sqrt{k} = 2\sqrt{\ra \rb} \le \ra + \rb.
\]

Since this is true and $k$ is nondecreasing (and usually strictly increasing) after each trade, then the total number of coins in Uniswap can never decrease below the twice the square root of the initial product by only trading between coins $\alpha$ and $\beta$.

\paragraph{Increasing liquidity with increasing reserves.}
Intuitively, the larger the amount of reserves, the less any one trade will cost. The marginal cost change of Uniswap (the negative of the infinitesimal price change of the Uniswap market after an infinitesimal trade) can be computed by differentiating~\eqref{eq:const-prod} twice and is given by
\begin{equation}\label{eq:slippage}
\frac{d^2\db}{d \da^2}\bigg|_{\da = 0} = \frac{2m_u}{\gamma \ra},
\end{equation}
where $m_u = \rb/\ra$ is the Uniswap price without fees. Note that~\eqref{eq:slippage} is strictly decreasing as $\rb$ increases (assuming $m_u$, the Uniswap price, stays constant).

We can similarly show this property directly by assuming that we have two markets, one with strictly larger reserves $\ra' > \ra$, and both with price
\[
\frac{\rb}{\ra} = m_u = \frac{\rb'}{\ra'}.
\]
In this case, assuming we trade $\da$ coins with both markets, we can show that the markets will have a price gap of
\begin{equation}\label{eq:asymptotic}
\db' - \db = m_u \gamma^{-1} \da^2\left(\frac{1}{\ra} - \frac{1}{\ra'}\right) + O\left(\frac{\da^2}{\ra^2}\right).
\end{equation}
This price gap is always positive in the case that $\ra' > \ra$ and this construction gives essentially a more precise version of~\eqref{eq:slippage}. The derivation can be found in appendix~\ref{app:price-gap}.

\paragraph{Cost of manipulation.} In the no-fee case with an infinitely liquid reference market, the cost of manipulating the price of a constant product market can be bounded from below.

If the attacker wishes to manipulate the constant product market to be $(1+\eps)m_p$, where $m_p$ is the true market price and $\eps > 0$ is some desired constant, then the attacker requires an amount of, at least
\begin{equation}\label{eq:manipulation-cost}
C(\eps) \ge K\rb \min\{\eps^2, \sqrt{\eps}\},
\end{equation}
coin $\beta$ for each period (here, a period could be, \eg, the time taken for a block to be mined) with $K > 0$ a universal constant.

This result implies that the cost of manipulation scales linearly with the reserve amounts, marking the importance of having large reserve pools. Additionally, the result extends immediately (as a lower bound) to the case with fees. For a derivation of~\eqref{eq:manipulation-cost} and an explicit bound on $K$, see appendix~\ref{sec:cost-manipulation}.

\paragraph{Liquidity provider returns.} In the no fee case (\ie, when $\gamma = 1$), we can easily compute the portfolio value of the Uniswap contract (and, correspondingly, any liquidity provider). Let $\ra^t$, $\rb^t$, and $m_p^t \in \reals_+$ be the reserves for coin $\alpha$, coin $\beta$, and the market price of coin $\alpha$, respectively, at each time $t=1, \dots, T$.

In the no fee case, the no-arbitrage bounds~\eqref{eq:no-arb} imply that, $m_p^t = \rb^t/\ra^t$, while the definition of a constant product market~\eqref{eq:const-prod} implies that $\ra^t\rb^t = k$ for all $t$. Combining both statements:
\[
\rb^t = \sqrt{k m_p^t}.
\]
We can use this expression to write a simple form for the relative return for Uniswap between time $t-1$ to $t$, given by
\[
\delta^t = \frac{m_p^t \ra^t + \rb^t}{m_p^{t-1} \ra^{t-1} + \rb^{t-1}} = \frac{\rb^t}{\rb^{t-1}} = \sqrt{\frac{m_p^t}{m_p^{t-1}}}.
\]
This implies that the total relative gain for a Uniswap contract is
\begin{equation}\label{eq:uniswap-returns}
\delta = \prod_{t=2}^T \delta_t = \sqrt{\frac{m_p^T}{m_p^1}},
\end{equation}
and the total portfolio value is
\begin{equation}\label{eq:uniswap-value}
P_V = (m_p^1\ra^1 + \rb^1)\delta = 2\sqrt{k m_p^T}.
\end{equation}
Since $\delta^t$ depends only on the relative ratios of $\ra$ and $\rb$, the result holds even when liquidity providers add tokens or remove tokens from the reserves---although in practice, as suggested by~\eqref{eq:slippage}, the price is likely to be less stable in the case of smaller reserves. See appendix~\ref{app:lp-returns-gbm} for further discussion.

\subsection{Discussion}
Because the arbitrage problem~\eqref{eq:main}, and its risky variant, problem~\eqref{eq:agent-arb}, are convex, we suspect that the no-arbitrage assumption is very likely to be met in practice. The convexity of this problem additionally implies that arbitrage can be efficiently performed even across several Uniswap markets (see appendix~\ref{app:Uniswap-convex} for more details).

As shown in~\eqref{eq:no-arb}, we analytically observe the effect that the Uniswap market fee has on the price of the market: the bounds guaranteed by the no-arbitrage assumption become looser. In other words, as the fee increases (or, equivalently, $\gamma$ decreases), we expect the Uniswap market to deviate further from the price of a given reference market.

Result~\eqref{eq:manipulation-cost} implies that, when the reserves are large, the cost of manipulation is generally expensive for all but the smallest of changes---though, due to the quadratic scaling when $\eps$ is small, we can expect small changes to the price to be relatively inexpensive. This implies that protocols which depend on Uniswap (or other constant product markets) as a price oracle should not be extremely sensitive to small price fluctuations; otherwise, it may be possible for an attacker with moderate resources to exploit this for their own gain.

These properties, combined with results~\eqref{eq:increasing-constant} and~\eqref{eq:asymptotic}, suggest that constant product markets are likely to be robust in the the practical setting where the number of tokens in reserve is large and the number of trades is also large.

\section{Constant mean markets}
A \emph{constant mean market} is a market which generalizes constant product markets. First introduced in~\cite{balancer}, constant mean markets satisfy the following equation in the absence of fees:
\begin{equation}\label{eq:const-mean}
\prod_{i=1}^n R_i^{w_i} = k,
\end{equation}
where $R_i$ are the reserves of coin $i = 1, \dots, n$, $w \in \reals^n_+$ are the weights associated with each coin, and $k \in \reals_+$ is the constant product. In this case, the weights all satisfy $w \ge 0$ with $\ones^Tw = 1$. In other words, in the absence of fees, constant product markets ensure that the product of its reserves stays constant, while constant mean markets ensure that the weighted geometric mean of the reserves, $R_i$ for $i=1, \dots, n$, stays constant.

Similar to constant product markets, trading $\Delta_j$ amount of coin $j$ for some amount $\Lambda_{j\ell}$ of a distinct coin $\ell \ne j$ should always satisfy the equation,
\[
\left(\prod_{\substack{i=1\\i\ne j, \ell}}^n R_i^{w_i}\right)\left(R_j + \gamma_j\Delta_j\right)^{w_j}\left(R_\ell - \Lambda_{j\ell}\right)^{w_\ell} = k,
\]
where $(1-\gamma_j)$ is the percentage fee associated with trading coin $j$. The corresponding reserves, $R_j$ and $R_\ell$ are updated as in~\S\ref{sec:const-prod}, as is the mean constant, $k$.

Note that constant product markets are a special case of a constant mean market where $n = 2$ and we have $w_1 = w_2 = 1/2$ and $\gamma_1 = \gamma_2$, with the constant product $k$ of~\eqref{eq:const-prod} replaced with its square root.

\subsection{Optimal arbitrage problem}
We can write the optimal arbitrage problem for constant mean markets in the following way:
\begin{equation} \label{eq:const-mean-opt}
\begin{aligned}
& \text{maximize} & & \sum_{i=1}^n\left(\sum_{j=1}^n \Lambda_{ij}m^p_j - \Delta_im_i^p\right)\\
& \text{subject to} & & \prod_{i=1}^n \left(R_i + \gamma_i \Delta_i - \sum_{j=1}^n \Lambda_{ij}\right)^{w_i}=k\\
&&& \Delta_i \ge 0, \quad i=1, \dots, n\\
&&& \Lambda_{ij} \ge 0, \quad i, j =1, \dots, n,\\
\end{aligned}
\end{equation}
where we will assume that $\Lambda_{ii}$ is constrained to be zero for notational convenience.\footnote{If $\gamma_i < 1$, it is not hard to prove that any locally optimal point will have $\Lambda_{ii} = 0$.} The variables in this optimization problem are the amount $\Delta \in \reals^n$ whose entries state how much of each coin to purchase from the external market, while $i,j$th entry of $\Lambda \in \reals^{n \times n}$ states how much of coin $i$ to trade for coin $j$.

Since the weighted geometric mean is a concave function that is increasing in all of its arguments~\cite[\S3.1.5]{cvxbook}, it turns out that the equality constraint in problem~\eqref{eq:const-mean-opt} can be relaxed to an inequality constraint to get an equivalent, but convex, optimal arbitrage problem~\cite[\S3.2.4]{cvxbook}. See appendix~\ref{app:mean-convex} for more details.

\subsection{Extensions and properties}
Surprisingly, almost all conditions and properties (except the no-arbitrage condition of~\eqref{eq:no-arb}) that hold for constant product markets also hold for constant mean markets in a similar form. We describe a few cases below.

\paragraph{Extensions.}
All of the same extensions given in section~\S\ref{sec:extensions} hold for problem~\eqref{eq:const-mean-opt}. More specifically, for any convex function $f:\reals_+^{n^2} \times \reals_+^{n} \to \reals$, which is increasing in its first $n^2$ arguments and decreasing in the remaining $n$ arguments, the following convex relaxation of~\eqref{eq:const-mean-opt} is an equivalent (but convex) problem with the additional penalty term, $f$:
\begin{equation} \label{eq:const-mean-cost}
\begin{aligned}
& \text{maximize} & & \sum_{i=1}^n\left(\sum_{j=1}^n \Lambda_{ij}m^p_j - \Delta_im_i^p\right) - f(\Lambda, \Delta)\\
& \text{subject to} & & \prod_{i=1}^n \left(R_i + \gamma_i \Delta_i - \sum_{j=1}^n \Lambda_{ij}\right)^{w_i} \ge k\\
&&& \Delta_i \ge 0, \quad i=1, \dots, n\\
&&& \Lambda_{ij} \ge 0, \quad i, j =1, \dots, n,\\
\end{aligned}
\end{equation}
with the same variables, $\Delta \in \reals^n$ and $\Lambda \in \reals^{n \times n}$, as problem~\eqref{eq:const-mean-opt}.

\paragraph{Properties.} Additionally, some of the same properties given in~\S\ref{sec:other-properties}, hold essentially in their exact form for constant mean markets: all instances will have an nondecreasing product constant and satisfy the corresponding no-depletion property. The notion of increasing liquidity with increasing reserves also holds and can be easily derived from~\cite[\emph{Out-Given-In}]{balancer}.

\paragraph{No-arbitrage conditions.}
It is, in general, not clear that there exists a closed-form solution for the no-arbitrage conditions specified in~\S\ref{sec:opt-arb}. It is possible to give simple necessary (but not sufficient) conditions on each pair of coins found in a given constant-product market via a similar argument to the one in~\S\ref{sec:opt-arb}, but we expect the general conditions are more complicated.

\subsection{Discussion}
We present these results for constant mean markets as they carry over nearly immediately from those given in~\S\ref{sec:const-prod}. Though we suspect that markets which satisfy the constant mean property, such as Balancer~\cite{balancer}---where the idea originated---may become more important in the near future, we focus on the specific case of Uniswap (and, more generally, constant product markets) as other protocols such as Celo~\cite{celo} heavily depend on the robustness of this particular market maker mechanism.

\section{Agent-based simulation of Uniswap markets}
While the properties presented in~\S\ref{sec:const-prod} lead us to believe that the Uniswap market is likely well-behaved under most scenarios, it is hard to make stronger claims about the robustness of the Uniswap market mechanism without making assumptions that are unlikely to be realistic.

To verify this experimentally, we created an agent-based simulation by using the Gauntlet DSL to specify how several types of agents interact with the current Uniswap contract\footnote{As of this time, the tested contract was based on commit \texttt{c10c08d} in the Uniswap Github repository, \texttt{https://github.com/Uniswap/contracts-vyper}.} on a simulated Ethereum blockchain. The simulation environment interacts with the Uniswap contract deployed on a simulated blockchain via Python bindings. The environment allows for configuration of the network's initial conditions, including distributions of agent behaviors and agent-specific parameters.

The simulation is run for a pre-defined number of time steps. For each simulated time step, environment state variables are updated based on the state of the on-chain contracts. We additionally evaluate policies for adding new agents to the environment. We evaluate the utility of an action for a given agent, and execute agent actions that have positive utility by submitting the corresponding transaction to the blockchain.

\subsection{Simulation markets}
\label{sec:simulation-markets}
In the current simulation, we have two possible markets for agents to trade with: one is given by the Uniswap contract and the other is given by a simple stochastic market model.

\paragraph{Uniswap market.}
When interacting with the Uniswap contract, an agent has a few possible actions they can perform: the agent can either (a) trade coins $\alpha$ for $\beta$ (and vice versa) subject to the constant product market equation~\eqref{eq:const-prod}, or (b) add or remove liquidity.

In the latter case, an agent is able to add, say, $\db$ amount of coin $\beta$ to reserve $\rb$ and is required to also add $\ra \db/ \rb$ of coin $\alpha$ to reserve $\ra$. The agent is then awarded
\[
\duni = \frac{\db}{\rb}\runi,
\]
where $\runi$ is the total amount of outstanding UNI coins given by the contract. The agent can also similarly remove liquidity by burning $\duni$ coins. The contract then gives the agent
\[
\da = \frac{\duni}{\runi}\ra, \quad \db = \frac{\duni}{\runi}\rb,
\]
and burns the given $\duni$ coins.

This market mechanism allows the agents who purchase these UNI coins, often called liquidity providers, to earn a profit given by the exchange fee, assuming the market price stays constant. Additionally, it provides a mechanism for adding and removing to reserves, thus making the Uniswap market more liquid, as shown in section~\S\ref{sec:other-properties}.

For the remainder of this section, we will assume the price of a UNI token is exactly given by the market prices of the equivalent amount of coins $\alpha$ and $\beta$ that would be received by burning the token. Additionally, the Uniswap contract defines $\gamma = .997$, which is the value we use from here on out.

\paragraph{Reference market.}
The reference market follows a simple power law model where the price $m_p$ of some coin $\alpha$, is updated in the following way:
\[
m_p \mapsto m_p + \kappa\da^{1+\xi},
\]
where $\kappa \ge 0$ and $\xi \ge 0$ are given in the problem data.
While it is possible for an arbitrageur to solve the arbitrage problem exactly (as it is a special case of~\eqref{eq:agent-arb}), we choose a simpler risk model for the arbitrageur as the true market price model is not known in practice.

We additionally update the market price every time step (after all agents have completed their actions) in the following way:
\[
m_p \mapsto m_p\cdot e^{\sigma X + \mu}.
\]
Here $X \sim \mathcal{N}(0,1)$ is drawn from a normal distribution and $\mu, \sigma \in \reals$ represent the mean returns and volatility of the market when no trades are performed.

\subsection{Simulation agents}
\label{sec:agents}
There are three broad classes of agents used in this simulation: arbitrageurs (who attempt to profit from deviations between Uniswap and the market), liquidity providers (who hold portfolios of UNI coins and currencies $\alpha$ and $\beta$), and traders (who trade coins with the Uniswap market, subject to simple rules). We describe each type of agent in detail below.

\paragraph{Arbitrageurs.}
Arbitrageurs seek to maximize their profit by trading between the Uniswap market and a reference market. An arbitrageur agent solves an instance of problem~\eqref{eq:agent-arb} with a simple quadratic cost of risk model:
\[
f_\alpha(\da) = \frac{\rho_\alpha}{2} \da^2, \quad f_\beta(\db) = \frac{\rho_\beta}{2} \db^2,
\]
and performs the necessary trades. Here, $\rho_\alpha, \rho_\beta \ge 0$ are parameters which control the penalty incurred for a trade.

We use this model instead of the riskless case of problem~\eqref{eq:main}, as live systems have noise during trades (\eg, network delays) such that opportunities found by exact minimization of the riskless problem~\eqref{eq:main} may not be easily executed or may close during the time it takes to perform the trade.

Additionally, we are able to model a wide variety of risk-taking behavior by arbitrageurs by, for example, increasing the parameters $\rho_\alpha$ and $\rho_\beta$ to have an arbitrageur who is less prone to perform risky (large) trades, even with large arbitrage opportunities available.

\paragraph{Liquidity providers.}
There are two types of liquidity providers we model in this simulation.

The first are the \emph{initial} liquidity providers, who begin the simulation by providing some amount of coins $\alpha$, and $\beta$ to the reserve and seek to gain profits by taking fees from Uniswap trades. We assume these liquidity providers will not withdraw their position until the end of the simulation.

The second are \emph{rational} liquidity providers. These agents perform Markowitz portfolio optimization (see~\cite[\S4.4.1]{cvxbook}) on the three possible coins in the market: $\alpha$, $\beta$, and the corresponding UNI coin minted from the Uniswap market. Each agent then solves the (convex) portfolio optimization problem,
\begin{equation}
\label{eq:markowitz}
\begin{aligned}
& \underset{x}{\text{maximize}} & & \hat \mu^Tx - \frac\lambda 2 x^T\hat \Sigma x\\
& \text{subject to} & & 1^Tx = 1\\
&&& x \ge 0,
\end{aligned}
\end{equation}
and makes the appropriate trades to rebalance their portfolio. In this problem, $x \in \reals^3$ is the vector containing the portfolio positions of $\alpha$, $\beta$, and UNI respectively. The problem data $\hat \mu \in \reals^3$ is the vector of exponentially-weighted average returns for each of the three coins (with respect to the reference market), $\hat \Sigma \in \symm_+^3$ is the exponentially-weighted covariance of returns, and $\lambda > 0$ is the penalty incurred by the risk. Since~\eqref{eq:markowitz} is not known to have a closed form solution, we set up and solve problem~\eqref{eq:markowitz} using the CVXPY~\cite{cvxpy_rewriting} modeling language and the ECOS solver~\cite{ecos} in our simulation.

The rational liquidity providers simulate agents who seek to maximize profits by trading their positions in each coin and holding the respective coins to maximize their expected return on investment, based on observed historical averages.

\paragraph{Traders.}
The final agent we model is a trader. In this case, the trader seeks to trade some amount $\da$ of coin for some second amount $\db$ (or vice versa), so long as the price of trading $\da$ coins on Uniswap differs no more than a constant percentage off from the same trade in the reference market.

In a way, traders embody the `demand for liquidity,' or the fact that trades on Uniswap might happen due to exogenous influences. In our case, a trader will draw $\da$ (or $\db$) from some probability distribution and check if performing this trade in the Uniswap market is at most some percentage more expensive than performing it in the reference market. If not---\ie, if the agent is able to trade with Uniswap for a reasonable price---then the agent makes the trade using the Uniswap contract. Otherwise, no trade is performed.

\subsection{Results}
\label{sec:results}
The results of our simulations suggest that the theoretical results derived above hold in practice under a wide variety of market conditions.

\begin{figure}
    \centering
    \includegraphics[width=.7\linewidth]{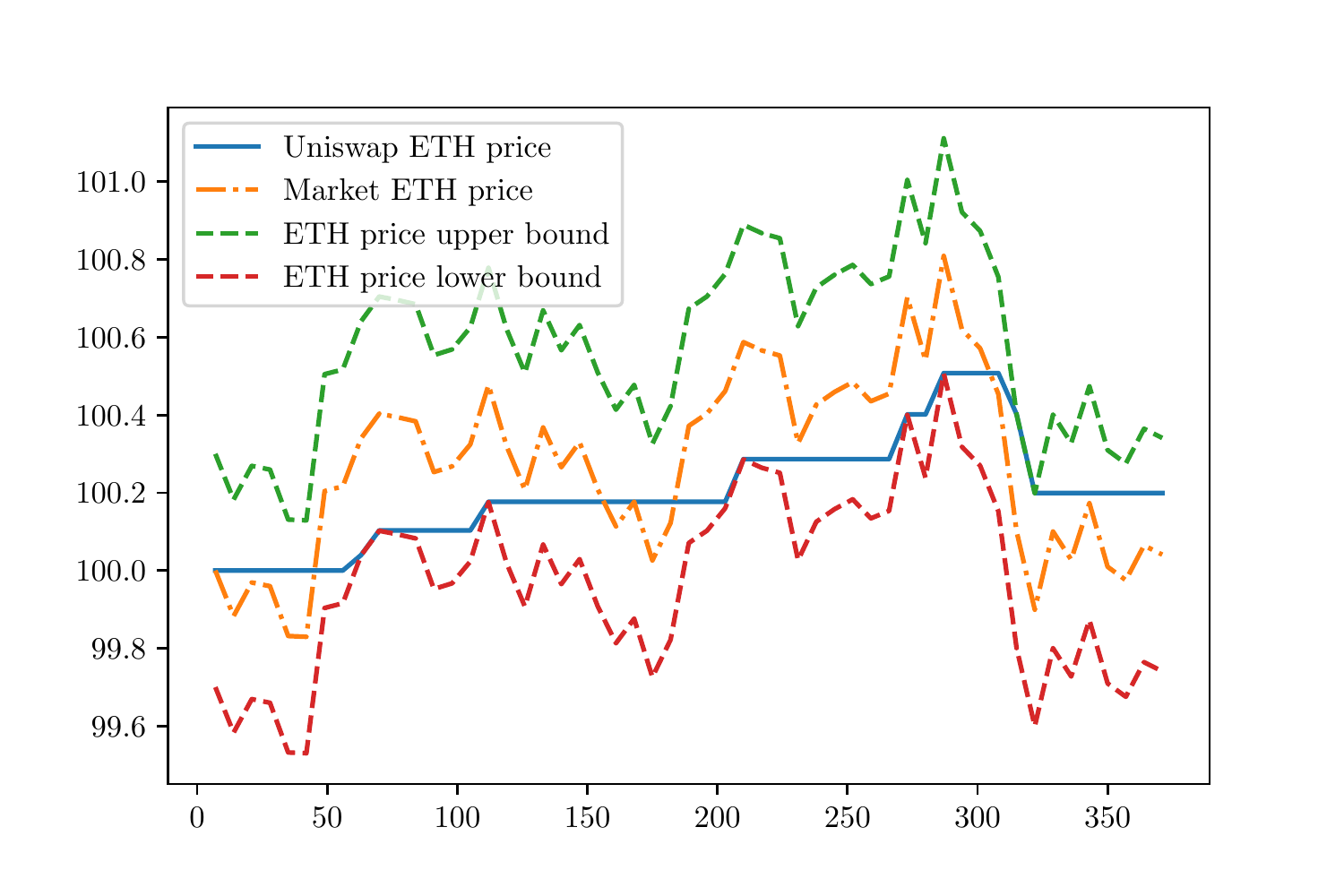}
    \caption{Market price ($m_p$) vs.\ Uniswap price ($m_u$) in no-drift condition with small noise and no traders. The plotted bounds are $\gamma m_p\le m_u \le \gamma^{-1}m_p$.}
    \label{fig:bounds-zoom}
\end{figure}

\begin{figure}
    \centering
    \includegraphics[width=.7\linewidth]{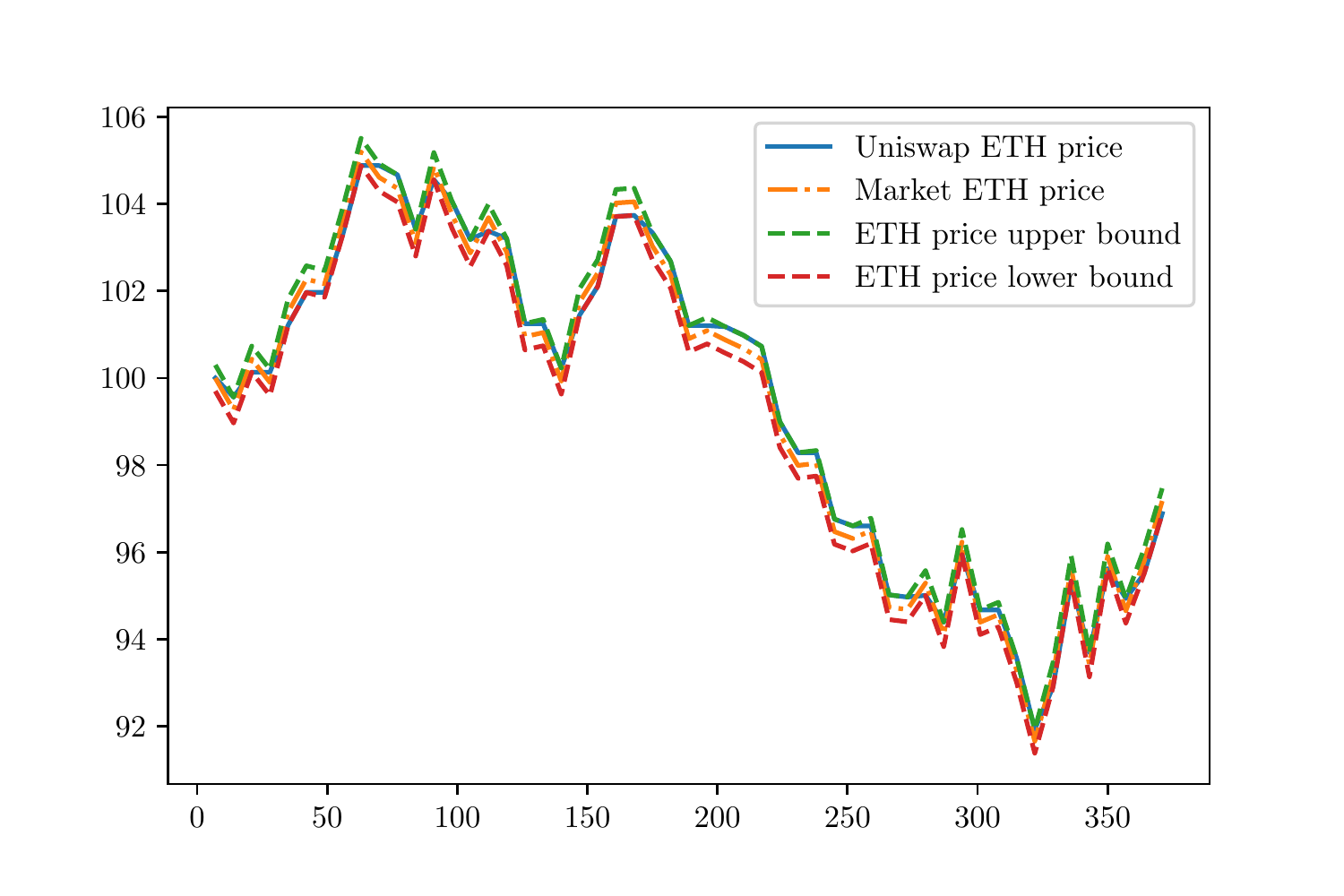}
    \caption{Market price ($m_p$) vs.\ Uniswap price ($m_u$) in negative-drift condition with moderate noise and no traders. The plotted bounds are $\gamma m_p\le m_u \le \gamma^{-1}m_p$}
    \label{fig:bounds-normal}
\end{figure}

\paragraph{Arbitrage bounds.} Figures~\ref{fig:bounds-zoom} and~\ref{fig:bounds-normal} show that, even under the presence of outside noise---such as random noise, or changes in reserves due to rational liquidity providers---the marginal Uniswap price stays within the predicted no-arbitrage bounds. In particular, figure~\ref{fig:bounds-zoom} shows when the market has a small amount of noise and little drift, while figure~\ref{fig:bounds-normal} shows this when there is large negative drift with moderate noise. We find that these bounds hold in our simulations, even under large amounts of market noise and large drift rates, so long as the trades performed are not large.

\begin{figure}
    \centering
    \includegraphics[width=.7\linewidth]{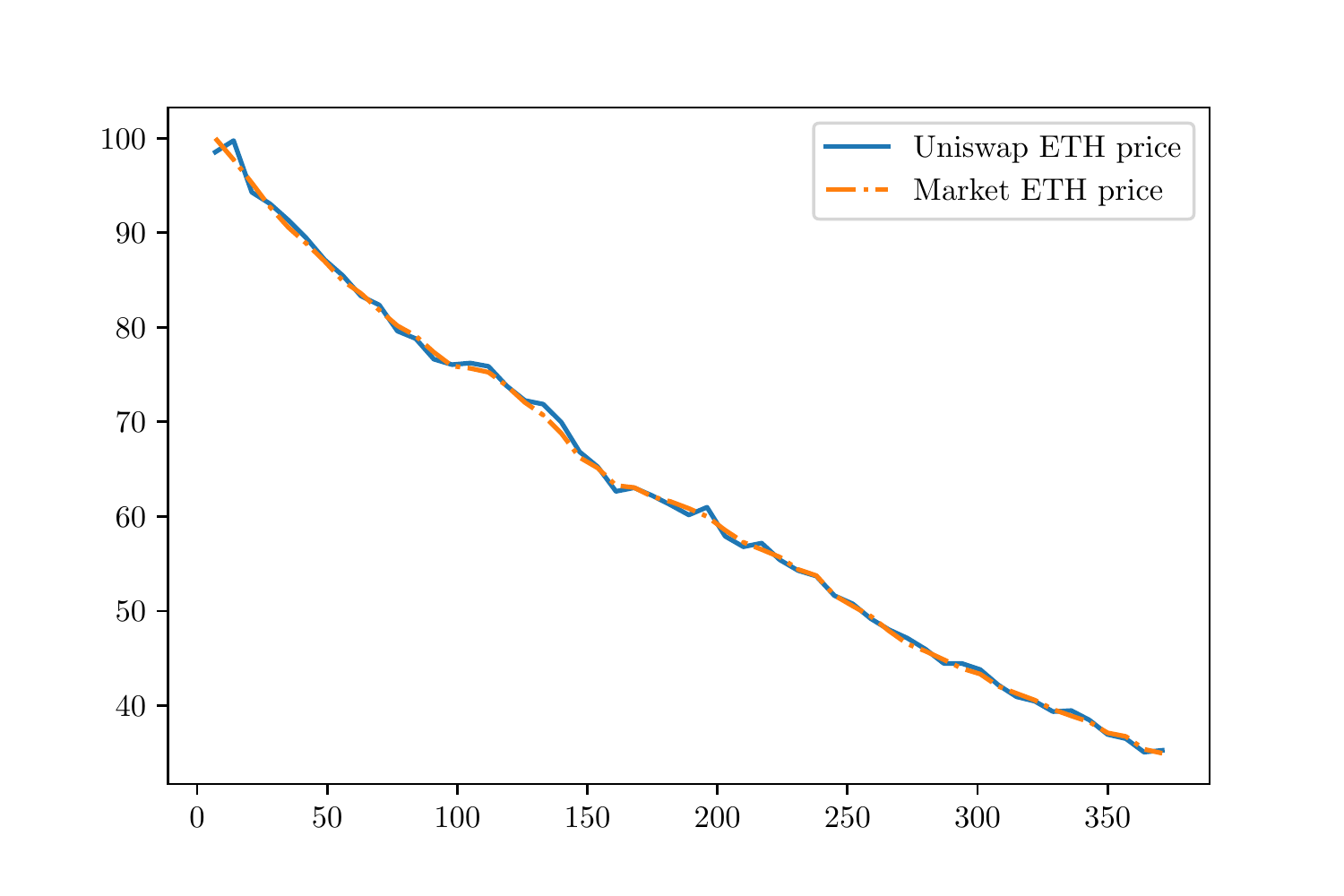}
    \caption{Market price ($m_p$) vs.\ Uniswap price ($m_u$) in large negative-drift condition with small noise, in the presence of traders. No bounds are plotted as they are nearly indistinguishable from the market ETH price, at this scale.}
    \label{fig:no-bound}
\end{figure}

Due to the discrete nature of the simulation, if a trader is the last agent to run at each time step, the no-arbitrage bounds could be broken as an arbitrageur may not have yet been able to perform arbitrage. While this is true (and is due to the nature of the simulation), the final result still clearly tracks the market price, as shown in figure~\ref{fig:no-bound}.

\paragraph{Initial liquidity provider returns.} In the simulation, we also record the utility of the initial liquidity providers, defined as the difference between the value of the respective UNI coin the agents hold, as defined in~\S\ref{sec:simulation-markets}, and a portfolio composed of the coins $\alpha$ and $\beta$ traded in for UNI at the start of the simulation.

\begin{figure}
    \centering
    \includegraphics[width=.7\linewidth]{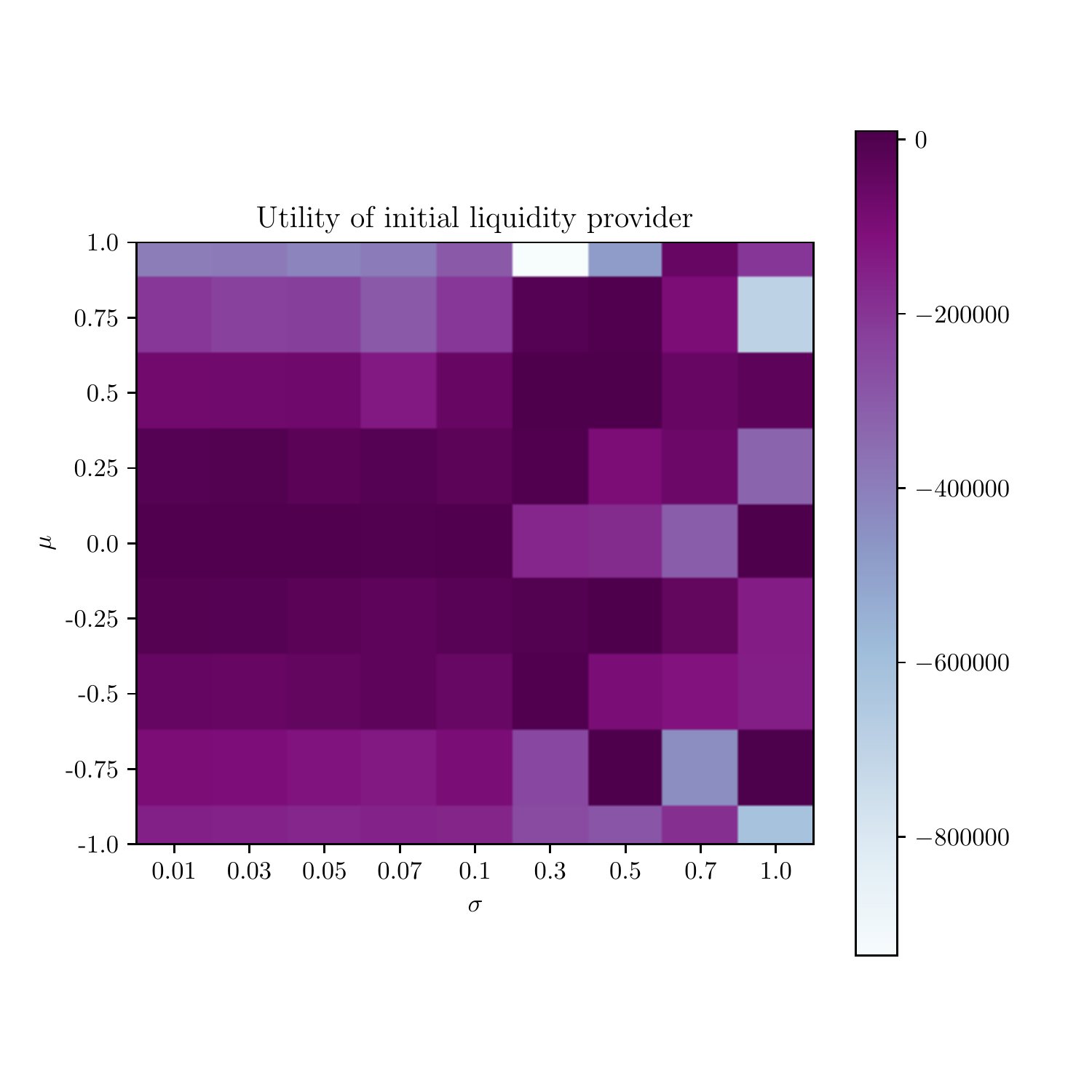}
    \caption{Varying utilities for initial liquidity providers over varying market conditions. As before, $\mu$ is the mean market return and $\sigma$ is the volatility.}
    \label{fig:initial-lp-utility}
\end{figure}

As shown in figure~\ref{fig:initial-lp-utility}, we find that in almost all of our simulations, initial liquidity providers end up having negative utility (using this definition), over most varying conditions. More specifically, we note that initial liquidity providers only have positive expected value when the mean market return is close to zero, as they earn revenue from all transaction fees while not falling behind an equivalent portfolio of the given pairs of coins. This approximately follows the result given in~\eqref{eq:uniswap-returns}, except that liquidity providers earn fees in this case.

\section{Conclusion}
Though simple, constant product markets and their generalizations have very nice theoretical properties (such as fairly strict no-arbitrage bounds on the reference price) which appear to hold in practice. Our simulations confirm that this is the case under a wide range of different market parameters and conditions, implying that the use of constant product markets as price oracles is, at least at first glance, sound.

Additionally, we suspect that there is an even larger class of automated market maker mechanisms which satisfy the above properties, and it would be interesting to further explore its mathematical properties. We leave this possible generalization for future work.

\section*{Acknowledgments}
The authors would like to thank Tony Salvatore for the initial implementation of the market model used in the simulation and Jonathan Tuck, Hayden Adams, Alex Evans, and Dan Robinson for helpful comments and edits on the paper. The authors would also like to thank Regina Cai, Georgios Konstantopoulos, and Merrick Wong for pointing out some inconsistencies in the text, reviewers 1 and 2 for their comments and suggestions, and Thoma Zoto for pointing out that the original proof in appendix~\ref{app:larger-incentives} was incorrect.


\newpage
\appendix
\section{The Uniswap arbitrage problem is convex}\label{app:Uniswap-convex}
We can easily show that this problem is convex by using~\eqref{eq:const-prod} to solve for $\db$,
\[
\db = \frac{1}{\gamma}\left(\frac{k}{\ra - \da} - \rb\right).
\]
Note that $\db$ is then a convex function of $\da$ as $x \mapsto 1/x$ is convex for $x$ positive, and a convex function composed with an affine function is convex~\cite[\S3.2.2]{cvxbook}. The resulting (equivalent) problem,
\begin{equation}\label{eq:Uniswap-cvx}
\begin{aligned}
& \text{maximize} & & m_p\da - \frac{1}{\gamma}\left(\frac{k}{\ra - \da} - \rb\right)\\
& \text{subject to} & & \da \ge 0,
\end{aligned}
\end{equation}
with variable $\da$ is then convex.

Since the problem is one-dimensional, the fact that problem~\eqref{eq:main} is convex is not immediately useful, but it can lead to several simple generalizations. For example, the optimal $m$-stage trading strategy with several interacting constant product markets can be efficiently evaluated in this case.

\paragraph{Optimality conditions.} Note that a maximum of a concave function over an interval happens either at (a) the interior of an interval or (b) at its boundary. In the latter case, it is not hard to show that a maximum is attained at the point on the boundary closest to the unconstrained maximum.\footnote{The proof follows from the fact that a concave function over $\reals$ is monotonically nondecreasing (nonincreasing) to the left (right) of its maximum.} Because of this, we only have to consider the unconstrained version of problem~\eqref{eq:Uniswap-cvx}, over the interval $[0, +\infty)$.

In this case, the unconstrained optimal points are those for which the objective of~\eqref{eq:Uniswap-cvx} has zero derivative. This happens when
\[
\da = \ra - \sqrt{\frac{k}{\gamma m_p}}.
\]
By the statement above, then the optimal solution to~\eqref{eq:Uniswap-cvx} is
\[
\da^\star = \left(\ra - \sqrt{\frac{k}{\gamma m_p}}\right)_+,
\]
where $(x)_+ = \max\{x, 0\}$ for $x \in \reals$. The optimal $\db^\star$ can be easily derived using the $\da^\star$ given above and the constant product formula~\eqref{eq:const-prod}.

Now, note that $\da^\star$ is zero if, and only if
\[
\ra - \sqrt{\frac{k}{\gamma m_p}} \le 0 \iff \gamma m_p \le  m_u,
\]
where $m_u = \rb/\ra$ is the marginal Uniswap market price of $\alpha$ without fees.

Since the objective of~\eqref{eq:Uniswap-cvx} is zero whenever there is no arbitrage opportunity (and the objective is only zero at $\da^\star = 0$, by strict convexity), then the above implies there is no $\alpha \to \beta$ arbitrage in the presence of an infinitely liquid market. Swapping $\alpha$ for $\beta$ yields the same result derived via no-arbitrage in the general case:
\[
\gamma m_p \le m_u \le \gamma^{-1}m_p.
\]

\subsection{Extensions to Uniswap arbitrage problem}
\label{app:Uniswap-extensions}
In a similar vein to~\eqref{eq:Uniswap-cvx}, we can add any penalty given by a convex function $f: \reals_+^2 \to \reals$ which is nondecreasing in its second argument to the objective. This yields yet another convex optimization problem:
\begin{equation}\label{eq:Uniswap-penalty-cvx}
\begin{aligned}
& \text{maximize} & & m_p\da - \db - f(\da, \db)\\
& \text{subject to} & & \frac{1}{\gamma}\left(\frac{k}{\ra - \da} - \rb\right) \le \db\\
&&& \da \ge 0,
\end{aligned}
\end{equation}
with variables $\da \in \reals$ and $\db \in \reals$. This problem is equivalent to problem~\eqref{eq:agent-arb} as the objective function is decreasing with respect to $\db$, implying that the first inequality constraint is always tight at an optimal point. Additionally, problem~\eqref{eq:Uniswap-penalty-cvx} is also convex as it is maximizing a concave function, subject to convex constraints.~\cite[\S4.2.1]{cvxbook}

\paragraph{Market models.} One particularly useful example of such a function $f$ is for modeling the market response of trading coin $\alpha$ for coin $\beta$. In particular, if the marginal market price $m:\reals_+ \to \reals_+$ is some decreasing function of the total amount of coin $\da$ traded, satisfying $m(0) = m_p$, then the slippage cost of trading $\da$ coin for $\db$ is given by,
\[
f(\da) = \int_0^{\da}(m_p - m(t))\,dt,
\]
with $f$ convex (as its derivative, $-m$, is increasing by definition). While several simple market models exist, the one we use in this paper is a special case of
\[
m(t) = m_p(1 - \eta t^{\xi}),
\]
with $\xi > 0$ and $\eta \ge 0$. This results in
\[
f(\da) = m_p\left(\frac{\eta}{\xi + 1} \da^{\xi+1}\right).
\]

\section{Derivation of price gap}
\label{app:price-gap}
By definition~\eqref{eq:const-prod}, the output of $\da$ can be written as
\[
\db = \frac{\rb\da}{\gamma(\rb - \da)}.
\]
Dividing the numerator and denominator by $\ra > 0$ and using the fact that $(1-x)^{-1} = 1 + x + O(x^2)$ we get the (nearly final) result:
\[
\db = \frac{\da m_u}{\gamma(1 - \frac{\da}{\ra})} =  m_u\gamma^{-1}\da\left(1 - \frac{\da}{\ra} + O\left(\frac{\da^2}{\ra^2}\right)\right).
\]
Subtracting $\db$ from $\db'$ and using the fact that $\ra < \ra'$ then yields the statement given in~\eqref{eq:asymptotic}.

\section{Uniswap portfolio value under Brownian dynamics}
\label{app:lp-returns-gbm}
Assume the trajectory of the market price $m_p^t$ follows a geometric Brownian motion process~\cite[\S5.1]{oksendalStochasticDifferentialEquations2013}
\[
\frac{{d}m_p^t}{m_p^t} = \mu dt + \sigma dW^{t},
\]
where $\mu \in \reals$ is the drift, $\sigma \in \reals_+$ is the volatility, and $W^{t}$ is a standard Brownian motion~\cite[\S3]{oksendalStochasticDifferentialEquations2013}. Without loss of generality, let $m_p^1 = 1$. The portfolio value then has expectation
\begin{equation}\label{eq:lp-portfolio-expect}
\Expect{[P_{V}]} = 2\sqrt{k}\Expect\left[\sqrt{m_p^t}\,\right] = 2e^{\frac{T}{8}\left( 4\mu{} - \sigma{}^2\right)}\sqrt{k} = 2e^{-\frac{T}{8}\sigma{}^2}\sqrt{k\Expect{[m_p^T]}},
\end{equation}
where we have used the fact that the $n$th moment of a log-normal random variable $X$ is given by~\cite{walck1996hand}
\[
\Expect\left[X^n\right] = e^{n\mu + n^2\sigma^2/2}.
\]

This argument shows one could replicate the time $T$ payoff of a no-fee liquidity provider with less initial capital than that required in the constant product market itself, which, in turn, implies that the portfolio value at time $T$ must grow at a rate of $e^{\frac{T}{8}\sigma^{2}}$ under the no-arbitrage hypothesis. The choice of $\gamma$ that gives this growth rate is the no-arbitrage fee. A simple replication of a volatility harvesting strategy can be used to show that some such $\gamma$ exists under mild conditions on $\mu$ and $\sigma$~\cite[\S11]{macleanKellyCapitalGrowth2011}.

\section{Splitting trades is always more expensive}
\label{app:larger-incentives}

Given two sequentially-feasible trades $(\da, \db)$ and $(\da', \db')$, we will show that the `aggregate trade' $(\da + \da', \db^\mathrm{tot})$ always satisfies $\db^\mathrm{tot} > \db + \db'$ for any fee $0 \le \gamma < 1$ with $\da, \da' > 0$. In
other words, the total output for trading $\da$ and $\da'$ sequentially is always smaller than the output of trading
$\da + \da'$ all at once.

To show this, note that after the trade $(\da, \db)$ is performed, the new reserves are given by $\ra + \da$ and $\rb - \db$. So,
since $(\da', \db')$ is a feasible trade when performed after $(\da, \db)$ (by definition) we have that
\[
(\ra + \da + \gamma\da')(\rb - \db - \db') = (\ra + \da)(\rb - \db),
\]
while, by definition, $(\da + \da', \db^\mathrm{tot})$ is feasible for reserves $\ra$, $\rb$, and we have
\[
(\ra + \gamma(\da + \da'))(\rb - \db^\mathrm{tot}) = \ra\rb.
\]
Since $(\da, \db)$ is also feasible for these reserves, we have that $(\ra + \gamma\da)(\rb - \db) = \ra\rb$ so,
\[
(\ra + \gamma(\da + \da'))(\rb - \db^\mathrm{tot}) = (\ra + \gamma\da)(\rb - \db). 
\]
Solving for $\db + \db'$, we find
\[
\db + \db' = \rb - \frac{(\ra + \da)(\rb - \db)}{\ra + \da + \gamma\da'},
\]
while solving for $\db^\mathrm{tot}$ gives
\[
\db^\mathrm{tot} = \rb - \frac{(\ra + \gamma\da)(\rb - \db)}{\ra + \gamma(\da + \da')}.
\]
This gives
\[
\db^\mathrm{tot} - (\db + \db') = (\rb - \db)\left(\frac{\ra + \da}{\ra + \da + \gamma\da'} - \frac{\ra + \gamma\da}{\ra + \gamma(\da + \da')}\right) > 0,
\]
where the inequality follows from the fact that $\rb - \db > 0$ combined with the inequality:
\[
\frac{x + z}{y + z} > \frac{x}{y},
\]
whenever $y > x \ge 0$ for any $z > 0$.

\section{Cost of manipulation}
\label{sec:cost-manipulation}
We can derive the cost of manipulation in the case where the reference market is infinitely liquid (\ie, when $\db = m_p \da$). We will derive this cost in the no-fee case as this lower bound on the cost is still a lower bound in the case with fees.

First, assume that an attacker wishes to increase the Uniswap price of a pair of coins $\alpha$ and $\beta$, by adding $\db$ coins to the system and removing $\da$ coins such that
\[
\frac{\rb + \db}{\ra - \da} = (1+ \eps)m_p,
\]
where $\eps > 0$ is some desired constant. Using the fact that $(\ra - \da)(\rb + \db) = k = m_p\ra^2$ (where the second equality follows from~\eqref{eq:no-arb} with $\gamma = 1$), we get that
\[
(1+\eps)m_p^2\ra^2 = (\rb + \db)^2,
\]
or, after some simplification,
\[
\db = \rb (\sqrt{1+\eps} - 1),
\]
required to increase the price. Since the attacker receives $\da = \rb(1 - (\sqrt{1+\eps})^{-1})/m_p$ as a result of this trade, the total cost of the attack is then
\[
C(\eps) = \db - m_p\da = \rb (\sqrt{1+\eps} + (\sqrt{1+\eps})^{-1} - 2),
\]
where $C(\eps)$ is the cost of carrying out the attack for a single time period.

\paragraph{Bounding from below.} Since $C(\eps)$ is twice differentiable in $\eps$ over $0 \le \eps \le 1$ with $C(0) = 0$ and is increasing, then,
\[
C(\eps) \ge \rb\frac{\eps^2}{2} \inf_{0 \le \eps' \le 1} C''(\eps) = \left(\frac{1}{32\sqrt2}\right)\rb\eps^2,
\]
where the second equality follows since $C''(\eps)$ is decreasing over $0 \le \eps \le 1$.

The $\eps \ge 1$ case is simple to bound by noting that
\[
x + x^{-1} - 2 \ge \kappa x,
\]
for all $x \ge \sqrt{2}$ whenever $\kappa = 3/2 - \sqrt{2}$ (this can be verified by multiplying the above inequality by $x$ to receive a convex quadratic whose largest root is $\sqrt{2}$). Therefore,
\[
\sqrt{1+\eps} + (\sqrt{1+\eps})^{-1} - 2 \ge \kappa \sqrt{1+\eps} \ge \kappa\sqrt{\eps},
\]
so the claim follows,
\[
C(\eps) \ge \kappa \rb\sqrt{\eps}.
\]
For an explicit bound on $K$, note that
\[
K \ge \min\left\{\frac{1}{32\sqrt{2}}, \frac{3}{2} - \sqrt{2}\right\} = \frac{1}{32\sqrt{2}},
\]
follows from the above.

The bounds, as stated, are tight up to a constant multiplicative factor (which follows from the asymptotic behavior of $C(\eps)$ as $\eps \downto 0$ and $\eps \upto + \infty$, respectively) although they are---in their current state---only useful as a theoretical tool, as the provided constants are very loose.

\section{The arbitrage problem in constant mean markets}
\label{app:mean-convex}
The problem
\begin{equation}
\label{eq:const-mean-cvx}
\begin{aligned}
& \text{maximize} & & \sum_{i=1}^n\left(\sum_{j=1}^n \Lambda_{ij}m^p_j - \Delta_im_i^p\right)\\
& \text{subject to} & & \prod_{i=1}^n \left(R_i + \gamma_i \Delta_i - \sum_{j=1}^n \Lambda_{ij}\right)^{w_i} \ge k\\
&&& \Delta_i \ge 0, \quad i=1, \dots, n\\
&&& \Lambda_{ij} \ge 0, \quad i, j =1, \dots, n,\\
\end{aligned}
\end{equation}
with variables $\Delta \in \reals^n$ and $\Lambda \in \reals^{n\times n}$ and with weights $w \ge 0$ and $\ones^Tw = 1$, is equivalent to problem~\eqref{eq:const-mean-opt} in that any optimal solution for problem~\eqref{eq:const-mean-cvx} is optimal for problem~\eqref{eq:const-mean-opt}.

Clearly, if any optimal point of~\eqref{eq:const-mean-cvx} has the first inequality constraint holding at equality, then this point is optimal for~\eqref{eq:const-mean-opt}. Because the objective is decreasing in $\Delta_i$ for each $i=1, \dots, n$ and increasing in $\Lambda_{ij}$ for each $i, j = 1, \dots, n$, while the geometric mean function in the inequality constraint is decreasing in $\Lambda_{ij}$ and increasing in $\Delta_i$, then the inequality constraint is always tight at equality at any optimal point.

Since the weighted geometric mean function is concave~\cite[\S3.1.5]{cvxbook}, then problem~\eqref{eq:const-mean-cvx} is a convex problem.

We also note that problem~\eqref{eq:const-mean-opt} is log-log convex~\cite{agrawalDisciplinedGeometricProgramming2019} with the obvious change of variables. While we do not explore this connection here, we suspect that this approach might provide a simple way of stating what class of functions can be used as an automated market maker mechanism such that the resulting no-arbitrage bounds are useful.

\newpage
\bibliographystyle{ieeetr}
\bibliography{bib}

\end{document}